\documentclass[aps,preprintnumbers,nofootinbib]{revtex4}
\usepackage{graphicx,epsfig,wrapfig}
\usepackage{amsmath,amssymb,bm}
\usepackage{slashed}

\usepackage{color}

\definecolor{darkgreen}{rgb}{0,0.65,0}

\newcommand{\be}{\begin{equation}}
\newcommand{\ee}{\end{equation}}
\newcommand{\ba}{\begin{eqnarray}}
\newcommand{\ea}{\end{eqnarray}}

\newcommand{\di}{\!{\rm d}}
\newcommand{\la}{\langle}
\newcommand{\ra}{\rangle}
\begin{document}
\title{\boldmath
	Monopole and quadrupole contributions to the angular momentum
	density}
\author{Peter Schweitzer}
	\affiliation{Department of Physics, University of Connecticut,
	Storrs, CT 06269, U.S.A.}
\author{Kemal Tezgin}
	\affiliation{Department of Physics, University of Connecticut,
	Storrs, CT 06269, U.S.A.}
\date{May 2019}
\begin{abstract}
The energy-momentum tensor form factors contain a wealth of information 
about the nucleon. It is insightful to visualize this information in terms 
of 3D or 2D densities related by Fourier transformations to the form factors. 
The densities associated with the angular momentum distribution were recently 
shown to receive monopole and quadrupole contributions. We show that these 
two contributions are uniquely related to each other. The quadrupole 
contribution can be viewed as induced by the monopole contribution, 
and contains no independent information. Both contributions however 
play important roles for the visualization of the angular momentum density.
\end{abstract}
%
%
%
\maketitle

\section{Introduction}
\label{Sec-1:introduction}

The form factors of the energy-momentum tensor (EMT) \cite{Kobzarev:1962wt} 
are a rich source of information on the structure of hadrons, whose
systematic exploration has begun only recently through studies of
generalized parton distribution functions \cite{Mueller:1998fv} 
entering the description of hard exclusive reactions, see
\cite{Ji:1998pc} for extensive reviews.

The 3D EMT densities were introduced in \cite{Polyakov:2002yz} as an 
important concept to visualize the information content of the EMT form 
factors in the nucleon. By considering Fourier transforms of the EMT 
form factors, one gains access to so far unexplored information ranging 
from the energy density, over angular and spin momentum densities, to 
mechanical properties of hadrons. A first visualization of the EMT 
densities based on calculations in the chiral quark soliton model was 
presented in \cite{Goeke:2007fp}. The EMT density formalism was further 
developed in \cite{Lorce:2017wkb,Polyakov:2018zvc}. 

In this note we focus on an important aspect of the interpretation of the
EMT form factor $J^a(t)$ where $a=g,\,u,\,d\,\dots$ denotes the parton species. 
In Ref.~\cite{Lorce:2017wkb} it was shown that the information content of 
the form factor $J^a(t)$ is described in terms of an angular momentum density 
which has a monopole contribution and a quadrupole contribution. 
The introduction of such densities (i) plays an important role in the 
visualization, and (ii) characterizes the independent nonperturbative 
information contained in form factors. Despite careful treatments in the 
Refs.~\cite{Polyakov:2002yz,Goeke:2007fp,Lorce:2017wkb,Polyakov:2018zvc},
these works remain incomplete with regard to the second aspect. 
The purpose of this work is to close this gap, and clarify what is the 
independent information contained in the 3D and 2D angular momentum 
densities of the nucleon.

For more aspects of EMT form factors regarding mechanical properties 
\cite{Polyakov:2018exb,Nature,Lorce:2018egm,Shanahan:2018nnv,Polyakov:2018rew}, 
the spin \cite{Leader:2013jra,Wakamatsu:2014zza,Liu:2015xha,Deur:2018roz}
and mass \cite{Ji:1994av,Lorce:2017xzd,Hatta:2018ina} decompositions, 
applications to charmonia 
\cite{Voloshin:1980zf,Novikov:1980fa,Sugiura:2017vks,Polyakov:2018aey}
and exotic hadrons \cite{Dubynskiy:2008mq,Eides:2015dtr,Perevalova:2016dln}, 
and extensions to higher spins 
\cite{Cosyn:2019aio,Polyakov:2019lbq,new-preprint}
we refer to the literature.

\section{EMT form factors and 3D densities}

The nucleon form factors 
(we use the notation of \cite{Lorce:2017wkb,Polyakov:2018zvc} with
$P= \frac12(p^\prime + p)$, $\Delta = p^\prime-p$, $t=\Delta^2$) of
the symmetric (Belifante-improved) EMT can be defined as
\ba
    \la p^\prime,s^\prime| \hat T_{\mu\nu}^a(0)|p,s\rangle
    = \bar u^\prime(p^\prime,s^\prime)\biggl[
      A^a(t)\,\frac{P_\mu P_\nu}{m}
    + J^a(t)\ \frac{i\,P_{\{\mu}\sigma_{\nu\}\rho}\Delta^\rho}{2m}
    + D^a(t)\,\frac{\Delta_\mu\Delta_\nu-g_{\mu\nu}\Delta^2}{4m}
    +{m}\,{\bar c}^a(t)g_{\mu\nu} \biggr]u(p,s)\,.
    \label{Eq:EMT-FFs-spin-12-alternative} \ea
The form factors of different partons $a=g,\,u,\,d\,\dots$ depend
on the (not indicated) renormalization scale, and satisfy 
$\sum_a A^a(0)=1$ and $\sum_a J^a(0)=\frac12$ reflecting that the 
EMT encodes information on the mass and the spin of the particle. The 
value of the $D$-term $\sum_a D^a(0)=D$ is not fixed \cite{Polyakov:1999gs}.
EMT conservation implies $\sum_a \bar{c}^a(t)=0\;\forall \,t$.

It is convenient to consider first the interpretation of EMT form 
factors in terms of 3D densities in the Breit frame characterized by 
$P=(E,0,0,0)$ and $\Delta=(0,\vec{\Delta})$ with $t=-\vec{\Delta}^2$ 
where one can introduce the static EMT \cite{Polyakov:2002yz}
\be\label{EQ:staticEMT}
	T^a_{\mu\nu}(\vec{r},\vec{s})=\int \frac{d^3 \Delta}{(2\pi)^3 2E}\ 
	e^{{-i} \vec{r}\vec{\Delta}} \langle p',s'| \hat{T}^a_{\mu\nu}(0)|p,s\rangle.
\ee
Here $\vec{s}$ denotes the polarization vector of the states 
$|p,s\ra$ and $|p^\prime,s^\prime\ra$ in their respective rest frames.
In this work we will focus on the Belifante-improved angular momentum 
density $J^{i,a}(\vec{r},\vec{s})=\epsilon^{ijk}r^jT^a_{0k}(\vec{r},\vec{s})$
\cite{Polyakov:2002yz}. 
In Ref.~\cite{Lorce:2017wkb} it was shown that this density has the following 
decomposition in terms of a monopole and a quadrupole contribution,
\be\label{Eq:def-AM}
	J^{i,a}(\vec{r},\vec{s})=
	J^{i,a}_{\rm mono}(\vec{r},\vec{s})+J^{i,a}_{\rm quad}(\vec{r},\vec{s})\, .
\ee
These densities correspond to $\la J^{i,a}_{Bel}\ra_{\rm mono}(\vec{r})$ 
and $\la J^{i,a}_{Bel}\ra_{\rm quad}(\vec{r})$ in the notation of 
Ref.~\cite{Lorce:2017wkb} and are defined as
\ba\label{Eq:def-mono}
	J^{i,a}_{\rm mono}(\vec{r},\vec{s}) &=& s^i
	\int\frac{\di^3\Delta}{(2\pi)^3}\,e^{-i\,\vec{\Delta}\,\vec{r}}\,
	\biggl[J^a(t) + \frac{2t}{3}\;\frac{\di J^a(t)}{\di t}\,
	\biggr]_{t=-\vec{\Delta}^2}\;, \\
   \label{Eq:def-quad}
	J^{i,a}_{\rm quad}(\vec{r},\vec{s}) &=& B_a^{ij}(\vec{r})\;s^j, \;\;\;\;\;
	B_a^{ij}(\vec{r})=
	\int\frac{\di^3\Delta}{(2\pi)^3}\,e^{-i\,\vec{\Delta}\,\vec{r}}\,
	\biggl(\Delta^i\Delta^j-\frac13\,\vec{\Delta}^2\,\delta^{ij}\biggr)
	\biggl[\;\frac{\di J^a(t)}{\di t}\,
	\biggr]_{t=-\vec{\Delta}^2} \; .
\ea
There is consensus in literature that the above decomposition is correct
\cite{Lorce:2017wkb,Polyakov:2018zvc}. The new insight is that the two 
densities $J^{i,a}_{\rm mono}(\vec{r},\vec{s})$ and 
$J^{i,a}_{\rm quad}(\vec{r},\vec{s})$ are not independent of each other
but characterized by one radial function $\rho_J^a(r)$ which has the 
property $\sum_a\int\di^3x\,\rho_J^a(r)=\frac12$ and encodes all independent 
information about the angular momentum density. 

\section{The monopole density}

The monopole contribution can be used to define the density $\rho_J^a(r)$ 
where $r=|\vec{r}|$ as
\be\label{Eq:def-mono-I}
	J^{i,a}_{\rm mono}(\vec{r},\vec{s}) = s^i\,\rho_J^a(r)\,,\quad
	\rho_J^a(r) = 
	\int\frac{\di^3\Delta}{(2\pi)^3}\,e^{-i\,\vec{\Delta}\,\vec{r}}\,
	\biggl[J^a(t) + \frac{2t}{3}\;\frac{\di J^a(t)}{\di t}\,
	\biggr]_{t=-\vec{\Delta}^2}\,.
\ee
Without loss of generality we choose the z-axis of the
$\vec{\Delta}$-integration to be along the vector $\vec{r}$, so
$\vec{\Delta}\,\vec{r}=\cos\theta\,r\,|\vec{\Delta}|$. Using the 
expansion of a plane wave $e^{-i\vec{\Delta}\,\vec{r}}$ in terms of spherical Bessel 
functions $j_l(x)=(-x)^l(\frac1x\,\frac{\di\,}{\di x})^l(\frac{\sin(x)}{x})$
and Legendre polynomials $P_l(x)$ and their orthogonality relation,
\be\label{Eq:expansion-plane-wave}
	e^{-i\vec{\Delta}\,\vec{r}} = \sum_{l=0}^\infty (-i)^l(2l+1)\,
	j_l(|\vec{\Delta}|r)\,P_l(\cos\theta)\;, \quad
	\int_{-1}^1\di\cos\theta\;P_l(\cos\theta)P_k(\cos\theta)
	=\frac{2}{2l+1}\,\delta_{lk}\,
\ee
we obtain from (\ref{Eq:def-mono-I}) the result
\be\label{Eq:def-mono-II}
	\rho_J^a(r) = 
	\int\frac{\di^3\Delta}{(2\pi)^3}\,j_0(|\vec{\Delta}|r)\,
	\biggl[J^a(t) + \frac{2t}{3}\;\frac{\di J^a(t)}{\di t}\,
	\biggr]_{t=-\vec{\Delta}^2} \,.
\ee
It is convenient to rename the dummy integration variable such that 
$|\vec{\Delta}|\to q$ and to express the derivative of $J^a(t)$ under
the integral of Eq.~(\ref{Eq:def-mono-II}) as
\be
	\biggl[\frac{\di J^a(t)}{\di t}\,\biggr]_{t=-q^2} 
	=	-\;\frac{1}{2q}\;\frac{\di J^a(-q^2)}{\di q} 
	\equiv  -\;\frac{1}{2q}\;\frac{\di J^a(q)}{\di q} 
\ee
where we in the last step we introduced the sloppy notation $J^a(t)\to J^a(q)$ 
to simplify the notation in the following.
We thus obtain
\be\label{Eq:def-mono-III}
	\rho_J^a(r) = 
	\int\frac{\di^3q}{(2\pi)^3}\,j_0(qr)\,
	\biggl[J^a(q) + \frac{q}{3}\;\frac{\di J^a(q)}{\di q}  \biggr] \,.
\ee

\section{The quadrupole density}

The quadrupole density is described by the $3\times3$ matrix $B_a^{ij}(\vec{r})$
which is symmetric and traceless. Notice that $\vec{r}$ is the only 
available vector in the integral defining $B_a^{ij}(\vec{r})$.
The symmetric matrix $B_a^{ij}(\vec{r})$ can therefore only be constructed 
from the tensors $\delta^{ij}$ and $r^i \, r^j$. On general grounds the 
matrix $B_a^{ij}(\vec{r})$ can be expressed as 
$B_a^{ij}(\vec{r})=\delta^{ij}\,a^a(r)+e_r^ie_r^j\,b^a(r)$
where $e_r^i= r^i/r$. Since $B_a^{ij}(\vec{r})$ is traceless, the functions 
$a^a(r)$ and $b^a(r)$ are actually not independent of each other, and satisfy
$B_a^{ii}(\vec{r})=3\,a^a(r)+b^a(r)=0$. Thus, the matrix $B_a^{ij}(\vec{r})$
is given by 
\be\label{Eq:matrix-B}
	B_a^{ij}(\vec{r}) = \biggl(\,e_r^i\,e_r^j-\frac13\,\delta^{ij}\biggr)\,
	b^a(r) \,.
\ee
In order to compute the function $b^a(r)$ we contract $B_a^{ij}(\vec{r})$ 
with the tensor $e_r^ie_r^j$ 
\ba
	e_r^i\,e_r^j\,B_a^{ij}(\vec{r})  = \frac23\,b^a(r) =
	\int\frac{\di^3\Delta}{(2\pi)^3}\,e^{-i\,\vec{\Delta}\,\vec{r}}\,
	\biggl((\vec{e}_r\vec{\Delta})^2-\frac13\;{\vec{\Delta}^2}\biggr)
	\biggl[\;\frac{\di J^a(t)}{\di t}\,
	\biggr]_{t=-\vec{\Delta}^2} \; .
\ea
Choosing the z-axis of the $\vec{\Delta}$-integration along the vector 
$\vec{r}$ we have $(\vec{e}_r\vec{\Delta})^2-\frac13{\vec{\Delta}^2}=
\frac23\,P_2(\cos\theta)\,\vec{\Delta}^2$ and exploring the plane
wave expansion and orthogonality of Legendre polynomials in
Eq.~(\ref{Eq:expansion-plane-wave}) we obtain
\ba\label{Eq:def-quad-II}
	b^a(r) = 
	\int\frac{\di^3\Delta}{(2\pi)^3}\;
	i^2\;j_2(|\vec{\Delta}|r)\;\vec{\Delta}^2
	\biggl[\;\frac{\di J^a(t)}{\di t}\,
	\biggr]_{t=-\vec{\Delta}^2}  = 
	\int\frac{\di^3q}{(2\pi)^3}\;
	j_2(qr)\,\frac{q}{2}\;\frac{\di J^a(q)}{\di q} \,.
\ea

\section{Proof that \boldmath $\rho_J^a(r)$ and $b^a(r)$ are related}

In order to prove that the densities $\rho_J^a(r)$ and $b^a(r)$ 
are related to each other, we notice that the integrand of $\rho_J^a(r)$ 
can be expressed as
\ba\label{Eq:Xa2}
	q^2\,j_0(qr)\biggl[J^a(q) 
	+ \frac{q}{3}\;\frac{\di J^a(q)}{\di q}\biggr]
	= - q^2\,j_2(qr)\;\biggl[\frac{q}{3}\;\frac{\di J^a(q)}{\di q}\biggr] +
	\frac1r\,\frac{\di\,}{\di q}\biggl[q^2\,j_1(q r)\,J^a(q)\biggr]\,,
\ea
which can be verified by using identities for spherical Bessel 
functions or by simply inserting their explicit definitions. 
The last term on the right-hand-side of Eq.~(\ref{Eq:Xa2}) is a total 
derivative in $q$ and drops out in the integral over $\,\di^3q$. Thus
we see from the identity (\ref{Eq:Xa2}) that the density $b^a(r)$ 
characterizing the quadrupole term can be expressed as
\be\label{Eq:relation}
	b^a(r) = -\frac{3}{2}\;\rho_J^a(r)\,,
\ee
and is therefore uniquely defined in terms of the monopole density. 

The relation of the monopole and quadrupole densities becomes most
lucid if we choose the nucleon polarization along a specific axis,
say z-axis. Both angular momentum densities have then only a z-component 
given by 
\ba\label{Eq:compact-3D}
	J^{z,a}_{\rm type}(\vec{r}) 
	= i^l\,P_l\biggl(\frac{z}{r}\biggr)\,\rho_J^a(r) 
	\quad\mbox{with}\quad
	\begin{cases} 	l=0 & \mbox{for} \quad  {\rm type} = {\rm mono},\\	
			l=2 & \mbox{for} \quad  {\rm type} = {\rm quad}.
	\end{cases}	
\ea

\section{Comment on Ref.~\cite{Goeke:2007fp}}

When defining the monopole density $\rho_J^a(r)$ we used the notation
of Ref.~\cite{Goeke:2007fp} where the density $\rho_J^a(r)$ was computed 
in the chiral quark soliton model for the flavor combination $Q=u+d$. 
What remains to be done is the proof that the  $\rho_J^a(r)$ defined in this 
work in fact coincides with the density introduced in Ref.~\cite{Goeke:2007fp}.

For that we invert the Fourier transform in Eq.~(\ref{Eq:def-mono}) 
and obtain
\be
	J^a(t)+\frac{2t}{3}\;\frac{\di J^a(t)}{\di t}
	= \int\di^3r\;j_0(r\sqrt{-t})\,\rho^a_J(r)
\ee
which is an ordinary linear differential equation for $J^a(t)$ with the 
initial condition $J^a(0)=\int\di^3r\;\rho^a_J(r)$.
The unique solution to this differential equation is 
\be
	J^a(t) = \int\di^3r\;\frac{3j_1(r\sqrt{-t})}{r\sqrt{-t}}\,\rho^a_J(r)
\ee
which coincides with the expression for $\rho_J(r)$ quoted 
in Eq.~(48) of Ref.~\cite{Goeke:2007fp}.

\section{Comment on 2D distributions}

The 3D density formalism is justified for heavy particles whose Compton 
wave length is much smaller than the particle size \cite{Sachs}.
This condition is very well satisfied for nuclei, and for the nucleon it 
is satisfied to a good approximation \cite{Hudson:2017xug}. The formalism 
of 2D lightcone densities has the advantage of being rigorous and free of 
approximations, even for light hadrons, as the transverse coordinates 
$\vec{b}_\perp$ remain invariant under boosts along the lightcone
\cite{Burkardt:2000za}. 

If we choose the z-axis as spatial direction for the lightcone the 2D
angular momentum densities can be derived (for type = mono, quad) 
from the 3D densities as \cite{Lorce:2017wkb}
\be
	J^{z,a}_{\rm type}(b_\perp) = 
	\int\limits_{-\infty}^\infty \di z\; J^{z,a}_{\rm type}(\vec{r})\,.
\ee
With the results from Eqs.~(\ref{Eq:compact-3D}) the 2D densities 
can be expressed as
\ba\label{Eq:compact-2D}
	J^{z,a}_{\rm type}(b_\perp) 
	= \int\limits_{-\infty}^\infty \di z\; i^l\,
	P_l\biggl(\frac{z}{\sqrt{b_\perp^2+z^2}}\biggr)
	\rho_J^a\biggl(\sqrt{b_\perp^2+z^2}\biggr) 
	\quad\mbox{with}\quad
	\begin{cases} 	l=0 & \mbox{for} \quad  {\rm type} = {\rm mono},\\	
			l=2 & \mbox{for} \quad  {\rm type} = {\rm quad}.
	\end{cases}	
\ea
We see that the monopole and quadrupole contributions are both
uniquely determined through integral relations in terms of the
same ``generating function'' $\rho^a_J(r)$. It is interesting to
remark that Eq.~(\ref{Eq:compact-2D}) could be used to define also
higher multipoles. The odd multipoles vanish (and are forbidden by
parity reversal in QCD). But even multipoles can be defined for all $l$.
Only the multipoles $l=0,\;2$ appear in the decomposition of angular
momentum densities. We are not aware whether higher even multipoles 
$l>2$ have a physical meaning.

\section{Visualization of the densities}

Let us assume for illustrative purposes that $J^a(t)$ has the following 
analytical form, which is a useful Ansatz for many form factors,
\be\label{Eq:J-ansatz}
	J^a(t) \stackrel{\rm Ansatz}{=} \frac{J^a(0)}{(1-t/M^2)^2} \,.
\ee
In this case the densities can be evaluated analytically, and we
find from Eqs.~(\ref{Eq:def-mono-III},~\ref{Eq:def-quad-II}) the results
\be\label{Eq:rhoJ-ansatz}
	\rho_J^a(r) =   J^a(0)\;\frac{M^4\!\!\!}{24\,\pi}\;r\;e^{-M\,r}\,,\quad
	b^a(r)      = - J^a(0)\;\frac{M^4\!\!\!}{16\,\pi}\;r\;e^{-M\,r}\,.\quad
\ee
The results in Eq.~(\ref{Eq:rhoJ-ansatz}) satisfy the general relation 
(\ref{Eq:relation}) as expected.

\begin{figure}[b!]
\centering

\vspace{-6mm}

\includegraphics[height=5cm]{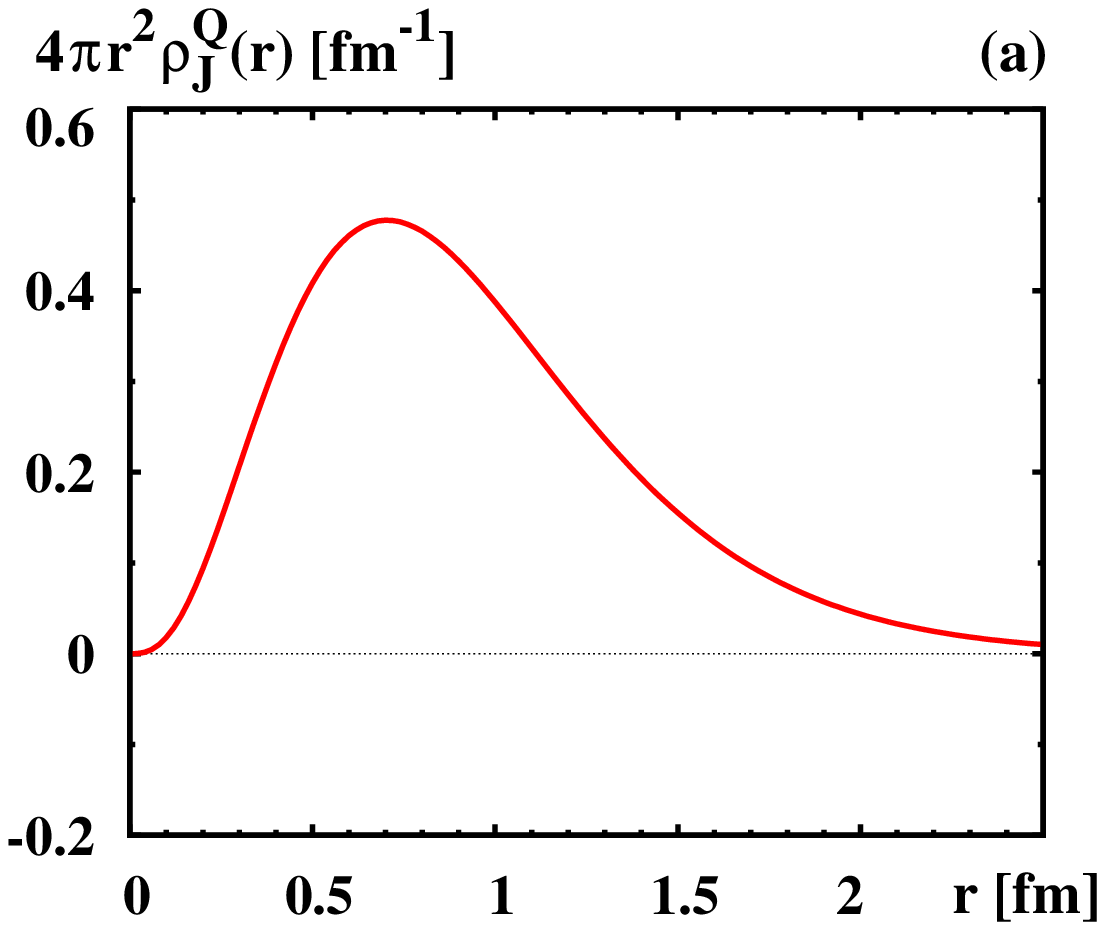} \ 
\includegraphics[height=5cm]{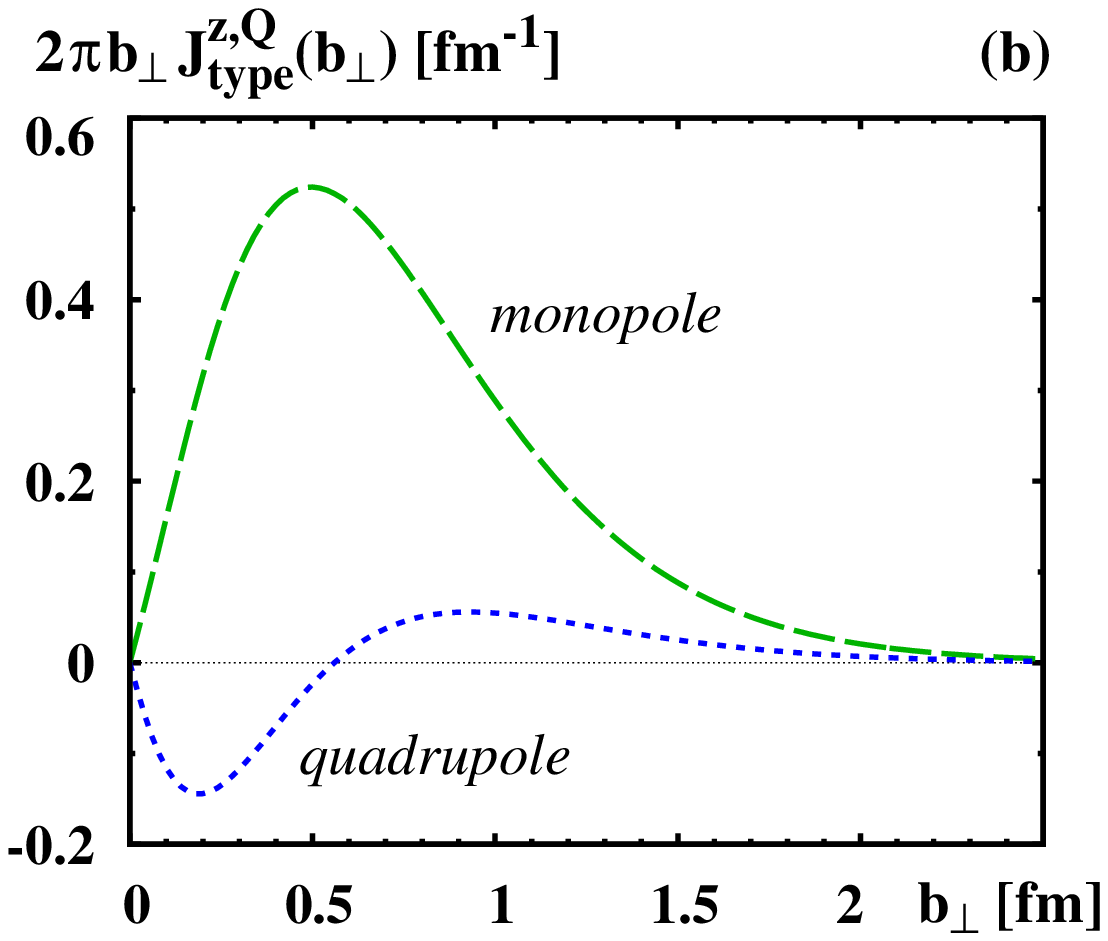} 

\vspace{6mm}

\caption{\label{Fig-1:densitities}
	(a) 
	3D Breit-frame density $\rho_J^Q(r)$ (solid line) which determines 
	the 3D monopole contribution to the angular momentum density via 
	Eq.~(\ref{Eq:def-mono-III}) and the 3D quadrupole contribution via
	Eqs.~(\ref{Eq:matrix-B},~\ref{Eq:relation}).
	(b) 
	The 2D lightcone densities of the monopole (dashed line)
	and quadrupole (dotted line) contributions, 
	$J_{\rm mono}^{z,Q}(b_\perp)$ and $J_{\rm quad}^{z,Q}(b_\perp)$,
	which are determined by means of Eq.~(\ref{Eq:compact-2D}).
	The densities satisfy $\int\di^3r\;\rho_J^Q(r)=\frac12$, 
	$\int\di^2b_\perp J_{\rm mono}^{z,Q}(b_\perp)=\frac12$ and
	$\int\di^2b_\perp J_{\rm quad}^{z,Q}(b_\perp)=0$.}
\end{figure}

In order to have a feeling how these densities look like, we use
results from the chiral quark soliton model \cite{Goeke:2007fp} 
which predicts $\la r_J^2\ra/\la r^2_{\rm ch}\ra\approx 1.5$ where 
$\la r_J^2\ra=\int\di^3r\,r^2\rho_J(r)/\int\di^3r\,\rho_J(r)$ is the 
mean square radius of the density $\rho_J(r)$ and $\la r_{\rm ch}^2\ra$ 
is the proton mean square radius defined analogously.
In this model the total form factor $J^Q(t)$, $Q=u+d$, can 
be approximated by the analytic expression (\ref{Eq:J-ansatz}). 
The numerical result for $\rho_J(r)$ from \cite{Goeke:2007fp} are 
reasonably approximated by the analytic form (\ref{Eq:rhoJ-ansatz}) 
in the range $0.3\lesssim r \lesssim 1.5\,{\rm fm}$ with 
$M\approx 0.83\,{\rm GeV}$. This is sufficient for our purposes 
to visualize the main features. The result for $\rho_J(r)$ from 
Eq.~(\ref{Eq:rhoJ-ansatz}) is shown in Fig.~\ref{Fig-1:densitities}a. 
The results for the 2D densities (\ref{Eq:compact-2D}) are generic, 
see Fig~\ref{Fig-1:densitities}b. 
Similar results were obtained for $J_{\rm mono}^{z,Q}(b_\perp)$ and 
$J_{\rm quad}^{z,Q}(b_\perp)$ in a scalar diquark model in 
Ref.~\cite{Lorce:2017wkb}. The main quantitative 
difference is that the results based on the chiral quark soliton,
Fig.~\ref{Fig-1:densitities}b, are much softer at small $b_\perp$
compared to the results from  Ref.~\cite{Lorce:2017wkb}. This is 
presumably due to the fact that the diquark model essentially 
describes the nucleon structure in terms of a hard perturbative 
nucleon-quark-diquark vertex, while the results from Ref.~\cite{Goeke:2007fp} 
are due to soft chiral interactions.

\section{Conclusions}

It was shown that the monopole and quadrupole contributions to the 
Breit-frame 3D angular momentum density of the Belifante-improved EMT 
are not independent of each other, but are characterized 
in terms of a density $\rho^a_J(r)$ normalized as
$\sum_a\int\di^3r\;\rho^a_J(r)=\frac12$. 
This due to the fact that the information content of one Lorentz-scalar 
form factor, like $J^a(t)$, is in one-to-one correspondence to one 3D 
density defined in the Breit frame, say $\rho^a_J(r)$. The polarization 
axis of the nucleon spin breaks spherical symmetry. This induces a 
quadrupole contribution which, however, contains no independent information, 
and is uniquely related to the monopole contribution. 
This is analog to the case of the mechanical densities, pressure $p(r)$
and shear forces $s(r)$, which are derived from the same form factor $D(t)$
and hence also not independent but related to each other by a differential 
equation following from EMT conservation \cite{Polyakov:2002yz}.

The monopole and induced quadrupole components are nevertheless both essential 
for the visualization of the angular momentum density $J^{i,a}(\vec{r},\vec{s})$ 
as a 3D vector field. The 2D monopole and quadrupole densities in elastic 
frames \cite{Lorce:2017wkb}, or equivalently on the lightcone in the 
Drell-Yan frame \cite{Burkardt:2000za,Lorce:2017wkb}, are expressed 
through integral relations in terms of $\rho_J^a(r)$.
In this work we focused on the Belifante-improved angular momentum density,
but the same result holds also for the monopole and quadrupole contributions
to several other 
densities defined in Ref.~\cite{Lorce:2017wkb}. 

This result is of importance for two reasons. First, it clarifies which 
information about the spatial distribution of the nucleon spin is independent, 
and which can be expressed in terms of other densities. Second, it is 
model-independent. This provides a valuable test and is worth exploring
in models
\cite{Wakamatsu:2006dy,Goeke:2007fq,Cebulla:2007ei,Grigoryan:2007vg,Pasquini:2007xz,Hwang:2007tb,Abidin:2008hn,Brodsky:2008pf,Burkardt:2008ua,Abidin:2009hr,Kim:2012ts,Jung:2014jja,Kanazawa:2014nha,Chakrabarti:2015lba,Mondal:2016xsm,Adhikari:2016dir,Kumar:2017dbf,Mondal:2017lph}, 
lattice QCD \cite{Mathur:1999uf,Hagler:2003jd,Gockeler:2003jfa,Hagler:2007xi,Engelhardt:2017miy,Bali:2018zgl}
and effective chiral theories \cite{Granados:2019zjw}.

\ \\
\noindent{\bf Acknowledgments.}
We are grateful to C.~Lorc\'e, L.~Mantovani, B.~Pasquini and
M.~V.~Polyakov for valuable discussions and comments on the
manuscript.
This work was supported by NSF grant no.\ 1812423 and
DOE grant no.\ DE-FG02-04ER41309.


\end{document}